\documentclass[aps,prl,twocolumn]{revtex4-1}

\usepackage{graphicx}

\usepackage{epsf}
\usepackage{amsmath}

\usepackage{hyperref}

\newcommand{\re}[1]{(\ref{#1})}

\newcommand{\up}{\uparrow}

\newcommand{\dn}{\downarrow}

\newcommand {\dis}{\displaystyle}

\newcommand{\beg}{\begin{equation}}
\newcommand{\en}{\end{equation}}

\newcommand{\eps}{\epsilon}
\newcommand{\lam}{\lambda}

\begin{document}

\title{Dynamics of emergent Cooper pairing at finite temperatures}
\author{Emil A. Yuzbashyan$^{1}$ and Oleksandr Tsyplyatyev$^{2,3}$}
\affiliation{$^1$Center for Materials Theory, Department of Physics and Astronomy,
Rutgers University, Piscataway, New Jersey 08854, USA \\
$^2$Physics Department, Lancaster University, Lancaster LA1 4YB, UK\\
 $^3$Department of
Physics, University of Basel, Klingelbergstrasse 82, CH-4056 Basel,
Switzerland}

\begin{abstract}
We study the time evolution of a system of fermions with pairing interactions at a finite temperature. The dynamics is triggered by an abrupt increase of the BCS coupling
constant. We show that if initially the fermions are in a normal phase, the amplitude of the BCS order parameter averaged over the Boltzman distribution of initial states exhibits damped oscillations
with a relatively short decay time. The latter is determined by the temperature, the single-particle level spacing, and the ground state value
of the BCS gap for the new coupling. In contrast, the decay is essentially absent when the system was in a superfluid phase before the coupling increase.
\end{abstract}

\maketitle

Considerable progress has been made over the past few years in understanding the dynamical fermionic pairing in response to  fast
perturbations[\citealp{Barankov2004}--\citealp{Emil2006}]. Recent interest in this
long-standing  problem[\citealp{Galaiko1972}--\citealp{Galperin}]  has been motivated by experiments
on cold atomic fermions with tunable interactions\cite{Regal2004,Zwerlein2004}, even though other systems have also been
considered\cite{Eastham,Papenkort}.

The general picture that emerged from the theory work is that as a result of the perturbation, e.g. a sudden change of the coupling
constant, the system of fermions with pairing interactions can reach a
variety of dynamical phases with properties quite distinct from the equilibrium ones [\citealp{Classify}--\citealp{Emil2006}]. For example, a steady
state characterized by undamped periodic oscillations of the time dependent Bardeen-Cooper-Schrieffer (BCS) order parameter $\Delta(t)$
[\citealp{Barankov2004},\citealp{Barankov2006}]
and a gapless steady state\cite{Barankov2006,Emil2006}, $\Delta(t) =0$,  have been identified.

Periodic oscillations occur in particular when at $t=0$ the fermions are described by a many-body wave function with a
seed
gap $\Delta_i$ much smaller than the ground state gap $\Delta_0$.   As a result of the Cooper instability of the initial state
the order parameter starts to grow exponentially, $\Delta(t)=\Delta_i e^{\Delta_0 t}$, and
reaches the ground state value at time $\tau/2=\ln(\Delta_0/\Delta_i)/\Delta_0$. However, in the absence of
the energy relaxation the system does not equilibrate and it can be shown that $|\Delta(t)|$
is  periodic in time  with a period $\tau$\cite{Barankov2004,Barankov2006}.

In this Letter we study the effect of temperature fluctuations on the non-adiabatic dynamics of fermions with attractive
interaction\cite{Barankovpra}. Suppose
 the system is initially in equilibrium at a finite temperature $T$. At $t=0$ the dynamics is triggered
 by an abrupt increase of the pairing strength and a certain quantity is measured at a later time. This process
 is repeated many times for each data point as is typical for measurements in atomic gases\cite{Zwerlein2005,Schunck}.
We are therefore interested in dynamical quantities averaged over the Boltzman distribution of initial states.

\begin{figure}[ht]
\includegraphics[width=3.in]{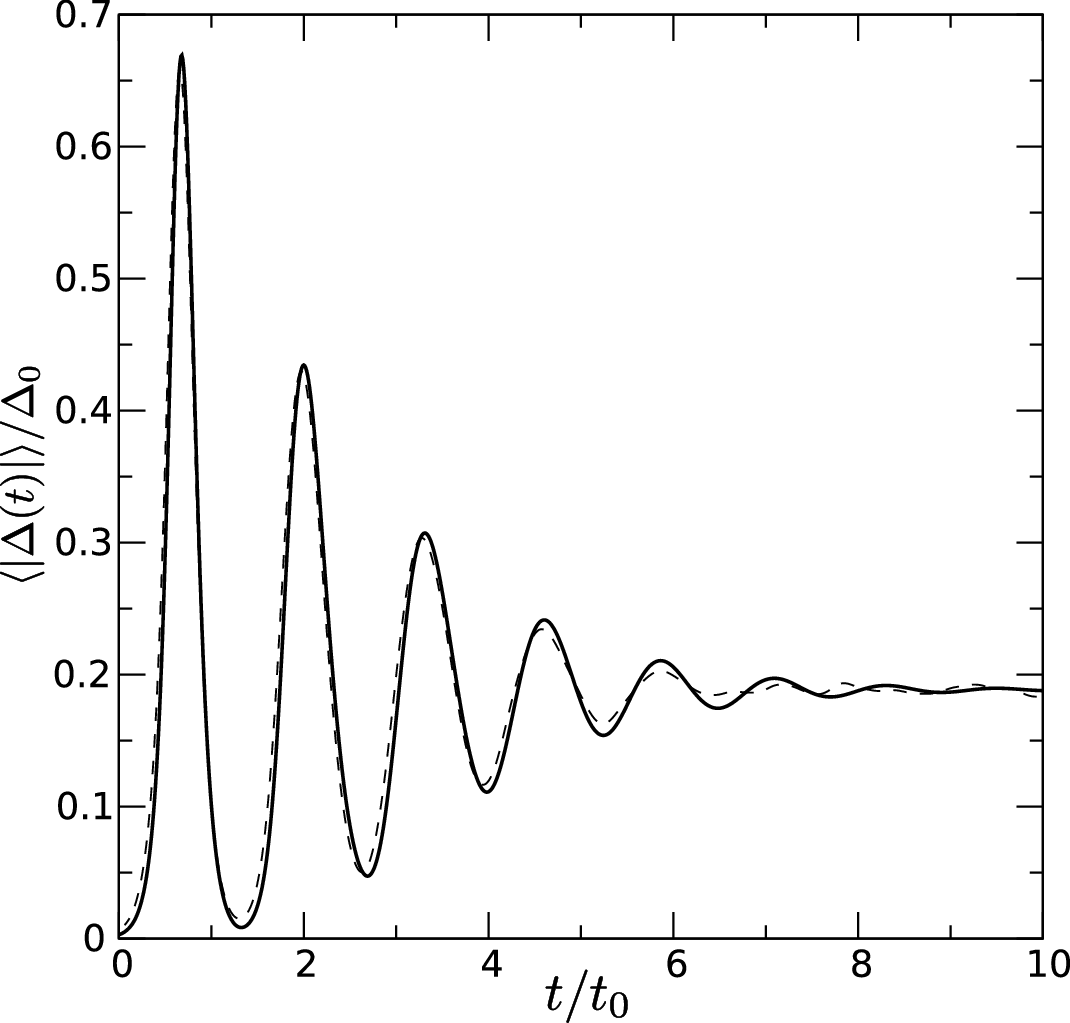}
\caption{Time evolution of the amplitude of the BCS order parameter $|\Delta(t)|$ averaged over initial states at temperature $T$.
 Numerical simulation of Eq.~\re{eqsm} for $10^4$ spins
averaged over $10^4$ realizations of initial conditions \re{blin} (solid curve) is compared to expression \re{avdef} (dotted curve).   The time is in units of the decay time $t_0$, Eq.~\re{taumain};  the ground state gap
is $\Delta_0=2\times 10^3\delta$ and $T=400\delta$, where $\delta$ is the level spacing. }
\label{fig}
\end{figure}

Our main results are as follows. We show that, if before the coupling increase the system is in a normal phase at temperature $T$, the average amplitude of the order parameter, $\langle|\Delta(t)|\rangle$, displays exponentially damped oscillations with a  decay time (see also Fig.~\ref{fig})
\beg
\frac{t_0}{\langle\tau\rangle} = \frac{1}{\pi^2} \ln\left( \frac{4\Delta_0^2}{T \delta }\right)
\label{taumain}
\en
where $\langle\tau\rangle$ is the average oscillation period and $\delta$ is the single particle level spacing. Here and below we assume
$\delta\ll T\ll\Delta_0$.     Expression \re{taumain} is accurate up to a prefactor of order one under the logarithm.

 For typical values for cold atomic fermions\cite{Regal2004,Zwerlein2004} Eq.~(\ref{taumain}) yields $t_0/\langle\tau\rangle\sim 1-3$,   i.e. there are only a few regular oscillations before the dephasing sets in.
In contrast, for the paired initial phase, we demonstrate that $t_0\propto 1/\sqrt{\delta}$ indicating that the decay time effectively diverges as the temperature is decreased below the critical temperature of the initial phase.

We emphasize that  each time the coupling is switched a particular initial condition is selected and the system goes into a state with periodic $|\Delta(t)|$. However,  whether the oscillations are seen in an
 ensemble averaged measurement  depends on the quantity being measured. For example,  it seems difficult to observe
 many of them in  $\langle|\Delta(t)|\rangle$. On the other hand, since the fluctuations of  the  oscillation frequency
 are small\cite{Barankovpra} (see also below Eq.~\re{period}), it can in principle be obtained e.g. from the ensemble averaged radio frequency absorption spectra\cite{Dzero2007}.

The decay time \re{taumain} can be qualitatively understood as follows. In the normal state  a nonzero initial value of
the order parameter $\Delta_i$ is due to fluctuations, which in mesoscopic samples are   governed by an energy
scale $\sqrt{T\delta}$ \cite{Scalapino,LarkinVarlamov}. Changing $\Delta_i$ by a factor of order one in the expression for the period $\Delta_0\tau=2\ln(\Delta_0/\Delta_i)$ leads to changes in the period  $\Delta_0\delta\tau\sim 1$. Then, one
expects the average of $|\Delta(t)|$ over all possible values of $\Delta_i$ to dephase  after $\tau/\delta\tau$ oscillations, i.e. on $t_0\propto \ln^2(\Delta_0/\sqrt{T\delta})/\Delta_0$ timescale. Note that
 the average period $\langle\tau\rangle\propto \ln(\Delta_0/\sqrt{T\delta})/\Delta_0$ and oscillation frequency
remain finite. In the superfluid state the order parameter has a macroscopic thermal average $\bar\Delta_i$, while typical thermal fluctuations $\sqrt{T\delta}\ll\bar\Delta_i$. In this case repeating the above argument, we obtain
$\Delta_0\delta\tau\sim \sqrt{T\delta}/\bar\Delta_i$ and  $\Delta_0 t_0\sim \bar\Delta_i\ln^2(\Delta_0/\bar\Delta_i)/\sqrt{T\delta}$, i.e. an extremely long decay time.

The non-stationary Cooper pairing   at times much shorter than the energy relaxation time can be described by the BCS model
\beg
\hat H={\sum_{j; \sigma=\dn, \up}\eps_{j} \hat c_{j \sigma }^\dagger \hat c_{j \sigma}-\lam\delta\sum_{j, k} \hat c_{j\up}^\dagger \hat c_{j\dn}^\dagger \hat c_{ k\dn} \hat c_{ k\up}},
\label{bcs1}
\en
where $\eps_j$ are the single fermion energies relative to the Fermi level,  $\delta$ is the mean spacing
between $\eps_j$, $\lam$ is the dimensionless BCS
coupling constant, and $\hat c_{j \sigma }$ are the fermionic annihilation operators.

In the time-dependent BCS mean-field
approach\cite{Barankov2004} the many-body wave function is a product state
\beg
|\Psi(t)\rangle=\prod_{n_m=0,2}\left(u_m(t)+v_m(t) c_{m\up}^\dagger \hat c_{m\dn}^\dagger\right)|0\rangle,
\label{wf}
\en
where $u_m(t)$ and $v_m(t)$ are the Bogoliubov amplitudes and the product is taken only over unoccupied ($n_m=0$)
and doubly occupied ($n_m=2$) levels. Singly occupied levels are excluded since their occupation numbers
are conserved by the Hamiltonian
\re{bcs1}.

The time evolution of the system is governed by the Bogoliubov-de Gennes equations
\begin{equation}
\begin{split}
&i\dot{u}_m=\epsilon_m u_m+\Delta v_m,
\quad i\dot{v}_m=-\epsilon_m v_m+\Delta^*u_m,
\end{split}
\label{bdg}
\end{equation}
where $\Delta=\lam\delta\sum_m u_m v_m^*$. These equations can be cast into the form of equations of motion
for classical spins\cite{Barankov2004}
\begin{equation}
\dot{\mathbf{s}}_{m}=2\mathbf{b}_m\times\mathbf{s}_{m},\quad\mathbf{b}_m=
\left(-\Delta_{x},-\Delta_{y},\epsilon_{m}\right),
\label{eqsm}
\end{equation}
where $\Delta_x$ and $-\Delta_y$ are the real and imaginary parts of $\Delta=\lam\delta\sum_m s_m^-$ and
the components of spins   are related to Bogoliubov amplitudes $u_m$ and $v_m$ as follows
\beg
2s_m^z= |v_m|^2-|u_m|^2,\phantom{l} s_m^-\equiv s_m^x- is_m^y=u_m v^*_m.
\label{bog}
\en
For example, according to Eqs.~(\ref{wf},\ref{bog})
the Fermi ground state where all states below the Fermi level are occupied and states above are empty
corresponds to $s_m^z=-\mbox{sgn } \eps_m/2$ and $s_m^-=0$.

Remarkably, nonlinear systems (\ref{bdg}) and \re{eqsm} turn out to be integrable\cite{Emil1}. The solution for $\Delta(t)$ can be obtained
with the help of the Lax vector technique \cite{Classify} by introducing
\beg
\mathbf{L}(w)=-\frac{\mathbf{z}}{\lam\delta}+\sum_{m}\frac{\mathbf{s}_{m}}{w-\epsilon_{m}},
\label{lax}
\en
where $w$ is an auxiliary parameter and ${\bf z}$ is a unit vector along the $z$ axis. The square of the Lax vector is
 conserved by Eq.~\re{eqsm} for any $w$ and therefore the roots of ${\bf L}^2(w)=0$ are
integrals of motion. Further, one can show that  the majority of the roots lie on continuous lines,
while the remaining isolated roots uniquely determine the form of $|\Delta(t)|$ at times $t\gg1/\Delta_0$\cite{Classify}. For instance, for initial
states close to the Fermi ground state there are two isolated
 roots, $w_1=i\gamma_1$ and $w_2=i\gamma_2$, in the upper half plane of complex $w$. In this case
the solution of Eq.~\re{eqsm} is known to be [\citealp{Barankov2004},\citealp{Emil1},\citealp{Barankovpra},\citealp{Classify},\citealp{Barankov2006}]
\beg
|\Delta(t)|=\Delta_+ \mbox{dn} \left(\Delta_+ (t-\tau/2), k\right),\quad k^2=1-\frac{\Delta_-^2}{\Delta_+^2},
\label{deltat}
\en
where $\mbox{dn}$ is the Jacobi elliptic function with modulus $k$, $\tau$ is its period, and $\Delta_\pm=|\gamma_1\pm\gamma_2|$.
Eq.~\re{deltat} describes periodic in time $|\Delta(t)|$ whose period and amplitude
are controlled by  $\Delta_\pm$.

First, consider a Fermi gas  at a temperature $T$ and zero  BCS coupling constant.
 At $t=0$ the coupling is suddenly turned on so that $\Delta_0\gg T$, where
$\Delta_0$ is the ground state gap for the new coupling. Before the interaction switch on the system can be
in any  eigenstate of the free Fermi gas
with the probability given by its Boltzman weight.  These states thus provide an ensemble of initial conditions
for equations of motion \re{eqsm} and our task is to evaluate the average of
$|\Delta(t)|$ over all possible initial states.

In the non-interacting problem amplitudes $(u_m, v_m)$ take values $(0,0)$, $(1,0)$, and $(0,1)$ corresponding to occupancies
$n_m=1$, 0, and 2, respectively.  Note that they
 are always correlated so that  $s_m^-=u_m  v_m^*=0$ and $s_m^z=\pm1/2$ or 0 indicating that the eigenstates of the free Fermi gas are (unstable)
 stationary states for the
 mean-field equations of motion (\ref{bdg},\ref{eqsm}). However, for any nonzero coupling they are not exact stationary states of the
  quantum Hamiltonian \re{bcs1} before the mean-field decoupling of the interaction term. These quantum effects facilitate the development of the Cooper instability and after
 a short time  states of the form \re{wf} with finite $u_m  v^*_m$ can be used. In the spin language, the spins
 ${\bf s}_m$ acquire nonzero $s_m^-$, i.e. nonzero components in the $xy$ plane.

As argued in Ref.~\onlinecite{Barankovpra} only spins at energies $|\eps_m|\lesssim T\ll\Delta_0$ initially have appreciable $xy$ components (see  below). It follows from
Eq.~\re{lax} that ${\bf L}^2(w)$ has two isolated roots in the upper half plane of complex $w$ and the order parameter
is described by Eq.~\re{deltat}, where the parameters $\Delta_\pm$ are
\beg
\Delta_+\approx \Delta_0,\quad \Delta_-\approx2\delta\biggl| \sum_m s_m^-\biggr|.
\label{pm}
\en
The values of $s_m^-$ are random with a distribution determined by the Boltzman distribution of initial states and the
quantum effects discussed above. On the other hand, there is a large number $N\sim T/\delta$ of random complex numbers in the sum \re{pm} and as noted in Ref.~\onlinecite{Barankovpra} (see Eq.~(46) therein)
 one therefore expects the Rayleigh distribution\cite{Barber}
\beg
p(\Delta_-)=C \Delta_- \exp\left( -\frac{\alpha\Delta_-^2}{4T\delta}\right)
\label{distg}
\en
independent of the details of the distribution of $s_m^-$. Here $\sqrt{T\delta}$ is a characteristic scale
of fluctuations of $\Delta_-$, $\alpha$ is of order one, and $C$ is a normalization constant.

Thus, averaging $|\Delta(t)|$ over Boltzman distributed initial states reduces to integrating Eq.~\re{deltat} with respect to
$\Delta_-$ with distribution \re{distg}, i.e.
\beg
\frac{\langle |\Delta(t)|\rangle}{\Delta_0} = \int_0^\infty \mbox{dn} \left(\Delta_0 (t-\tau/2), k\right)p(\Delta_-)d\Delta_- \label{avdef}
\en
Note that the Jacobi function dn depends on $\Delta_-$ through its modulus $k=1-\Delta_-^2/\Delta_0^2$.
For example, its period for $\Delta_-\ll\Delta_0$ is \cite{Abramowitz}
\beg
\tau=\frac{2}{\Delta_0}\ln\left(4\frac{\Delta_0}{\Delta_-}\right).
\label{per}
\en
 Using Eqs.~(\ref{per},\ref{distg}), we evaluate the average oscillation period and its standard deviation  (see also Ref.~\onlinecite{Barankovpra}),
\beg
\langle \tau\rangle=\frac{1}{\Delta_0}  \ln \frac{4\Delta_0^2}{T\delta}, \quad \delta\tau= \frac{\pi}{\sqrt{6}}\frac{1}{\Delta_0},
\label{period}
\en
up to a factor of order one under the logarithm. The average frequency and its deviation are $\langle\omega\rangle=2\pi/\langle \tau\rangle$ and $\delta\omega/\langle\omega\rangle=\delta\tau /\langle \tau\rangle$.

The asymptotic behavior of integral \re{avdef} at large times $t\gg t_0$ can be evaluated using the saddle point method
\beg
\begin{array}{ll}
\dis \frac{\langle |\Delta(t)| \rangle}{\Delta_0} =\frac{1}{\sqrt{\Delta_0 t_0}} - \frac{4 \sqrt{\Delta_0 t}}{\Delta_0 t_0} e^{-t/t_0} \cos \left[ \eta(t) \right],\\
\\
\dis \eta(t)=\frac{2 t}{\pi t_0}\ln \frac{2 t}{\pi t_0}- \frac{2t}{\pi t_0}+
\frac{2\Delta_0 t}{\sqrt{\Delta_0 t_0}}+
\frac{\pi}{4},\\
\end{array}
\label{deltaan}
\en
where $t_0$ is given by Eq.~\re{taumain}.   We see that on the $t_0$ time scale $\langle|\Delta(t)|\rangle$  exponentially approaches
a constant value smaller than the ground state gap $\Delta_0$ by a large factor $\ln(\alpha\Delta_0^2/T\delta)/\pi$. The approach is oscillatory with a typical period close to the ensemble averaged period
 $\langle\tau\rangle$.

Next, we present several alternative systematic derivations of Eq.~\re{distg} and show that it is independent of the details of initial state distribution. First, note that Eqs.~\re{eqsm} are equations of motion for classical spin Hamiltonian
$H=\sum_m 2\eps_m s_m^z-\lam\delta \sum_{m,n} s_m^+s_n^-$.
As discussed above,  before the interaction switch on  the spins are  along the $z$ axis. Their $z$ components take
values $s_m^z=\pm1/2$ or 0 with independent probabilities proportional to the corresponding Boltzman weight $e^{-2\eps_m s_m^z/T}$. This presents a technical difficulty, since these spin configurations
 are (unstable) equilibria for Eqs.~\re{eqsm}.

 One way to circumvent this problem is to replace the above ensemble of initial spin
 configurations with the Boltzman distribution of classical spins of length $s_m=1/2$.
 Then, each spin ${\bf s_m}$ is characterized by polar and azimuthal angles $\theta_m$ and $\phi_m$ with independent probability proportional to   $\exp(-\eps_m\cos\theta_m/T)$, i.e. spins at $|\eps_m|\lesssim T$ acquire finite components in the $xy$ plane. Using this probability distribution and Eq.~\re{pm}, we evaluate $p(\Delta_-)$. The calculation results in  Eq.~\re{distg} with  $\alpha=2/\ln(\Delta_0/T)$ and we obtain Eqs.~(\ref{taumain},\ref{period}) and \re{deltaan}.

 A distribution of the form \re{distg} for $|\Delta(t=0)|$ was  obtained in Ref.~\onlinecite{Barankovpra} for an ensemble of initial conditions suggested in the same reference.
 Note that according to Eq.~\re{wf} $|u_m|^2$ and $|v_m|^2$ represent probabilities of zero and double occupancy, respectively, of the level $\eps_m$.
In the free Fermi gas before the interaction is turned on their thermal averages are $\langle|u_m|^2\rangle=n_m^2$ and
 $\langle|v_m|^2\rangle=(1-n_m)^2$, where $n_m=(e^{\eps_m/T}+1)^{-1}$ is the Fermi function.
Averaging Eq.~\re{bdg} with respect to $\eps_m$ in
a narrow window of energies, we can replace $u_m$ and $v_m$
 with
$\langle u_m\rangle =e^{i\eps_m t} n_m^2$ and $\langle v_m\rangle =e^{-i\eps_m t} e^{i\phi_m} (1-n_m)^2$,
where $\phi_m$ is a random relative phase.   Since the total energy of the free gas does not depend on $\phi_m$, they are assumed to have independent uniform distributions.
Further, assuming
$\langle u_m  v^*_m\rangle\approx \langle u_m\rangle \langle v^*_m\rangle$, and using Eq.~\re{bog},
we obtain
the following initial spin configurations \cite{Barankovpra}
\beg
s_m^z=-\frac{1}{2}\tanh\left(\frac{\eps_m}{2T}\right),\quad s_m^-=\frac{e^{-i\phi_m}}{\dis4 \cosh^2\left(\frac{\eps_m}{2T}\right)}
\label{blin}
\en
Using Eqs.~(\ref{blin}, \ref{pm}) and uniform distributions for $\phi_m$, we  derive Eqs.~(\ref{taumain},\ref{period}) and Eq.~\re{distg} with $\alpha=6$, see also Fig.~\ref{fig}.

Finally, Eqs.~(\ref{taumain},\ref{period}) and \re{deltaan} can be derived starting from the Ginzburg-Landau  free energy. The advantage of
this approach is that we can consider initial states with nonzero BCS coupling that is suddenly increased at $t=0$.
The ground state gap  for the new coupling, $\Delta_0$, is
assumed to be much larger than that for the old coupling. Then, the equation ${\bf L}^2(w)=0$ has two isolated roots with
$\mbox{Im\,} w>0$  and the evolution of the order parameter is described by Eq.~\re{deltat} as before. With the help of Eq.~\re{lax}, we obtain
$\Delta_+\approx\Delta_0$ and $\Delta_-\approx2\Delta_i \ln (\Delta_0/\Delta_i)$, where $\Delta_i$ is the gap for the old coupling.

First, consider the case $T>T_c$, where $T_c$ is the critical temperature for the old coupling. To calculate the
average of $|\Delta(t)|$ over initial states, we need the probability distribution of $\Delta_-$ or equivalently the distribution
of possible values of the gap $\Delta_i$ before the coupling increase. We assume the latter is of the Ginzburg-Landau form
$\Delta_i \exp(-F(\Delta_i)/T)$, where the free energy for $T>T_c$ is
$F\left(\Delta_{i}\right)=\ln (T/T_c)|\Delta_{i}|^2/\delta$ \cite{Scalapino,LarkinVarlamov}. Using this distribution function and the above expressions for
$\Delta_\pm$ in terms of $\Delta_i$, we again obtain Eq.~\re{deltaan}, where now
\beg
t_0 \approx \frac{1}{\pi^2 \Delta_0} \ln^2\left( \frac{\ln(T/T_c)\Delta_0^2}{T \delta }\right),
\label{taugl}
\en
 This expression holds for $T-T_c\gg\sqrt{T_c\delta}$.

Below the critical temperature,  for $T_c> T_c-T\gg\sqrt{T_c\delta}$, we keep the quartic term in  $F(\Delta_i)$ and  expand Eq.~\re{avdef} in $\Delta_i- \bar\Delta_i$, where $\bar\Delta_i$ is the thermal
average of the order parameter before the coupling change. Using a saddle point method, we obtain a Gaussian decay to a
constant value on $t_0\propto  \bar\Delta_i^2\ln^2(\Delta_0/\bar\Delta_i)/\sqrt{T^3\delta}$ timescale. On the other hand, the dynamics at times
this long is likely not described by the Hamiltonian \re{bcs1} that does not account for energy relaxation.
Thus, we see that the dephasing of ensemble averaged oscillations due to thermal fluctuations is effectively absent when the dynamics is started in the paired phase. The reason is that in this case the order parameter has a macroscopic initial average much larger
than its thermal fluctuations. The fast dephasing  above $T_c$  crosses over into a slow dephasing below $T_c$ in a narrow window of temperatures
$|T-T_c|\sim\sqrt{T_c\delta}$.

In conclusion, we studied the effect of thermal fluctuations on the dynamics of fermions with pairing interactions triggered by an abrupt increase of
the  pairing strength. We showed that if the system is in the normal phase before the coupling increase, the amplitude of the order parameter averaged over the Boltzman distribution of initial states exhibits damped oscillations with relatively short decay time \re{taumain},  see Eq.~\re{deltaan}. On the other hand, the damping  is essentially absent when the dynamics starts from the superfluid phase.

An interesting  problem is to determine the time evolution described by the
{\it quantum} Hamiltonian \re{bcs1} at $T=0$ starting from the Fermi ground state, i.e. the ground state of the Hamiltonian \re{bcs1} for
$\lam=0$. Extending the above considerations to this case, one might expect damped oscillations  due
to quantum fluctuations {\it without} ensemble averaging. If this is the case, an estimate for the decay time  can be obtained
by replacing the temperature $T$ in Eq.~\re{taumain} with the level spacing $\delta$, i.e.  $t_0\sim \ln^2(\Delta_0/\delta)/\Delta_0$.

We thank B. L. Altshuler for many fruitful discussions. E.A.Y.`s research was supported
by a David and Lucille Packard Foundation Fellowship for Science and Engineering, NSF
award NSF-DMR-0547769, and Alfred P. Sloan Research Fellowship. O. T. acknowledges support from the EPSRC.

\end{document}